\documentclass[10pt,letterpaper]{article}
\usepackage{opex3}
\usepackage{color}
\usepackage{ulem}
\usepackage{amsmath}
\usepackage{upgreek}
\usepackage{float}
\usepackage[T1]{fontenc}
\usepackage[subrefformat=parens,labelformat=parens]{subfig}

\begin{document}

%
%

\title{Detection of non-classical space-time correlations with a novel type of single-photon camera}

\author{Felix Just$^{1}$, Mykhaylo Filipenko$^{2}$, Andrea Cavanna$^{1}$, Thilo Michel$^2$, Thomas Gleixner$^2$, Michael Taheri$^1$, John Vallerga$^3$, Michael Campbell$^4$, Timo Tick$^4$, Gisela Anton$^2$, Maria V. Chekhova$^{1,5,6}$, and Gerd Leuchs$^{1,6}$}

\address{$^1$ Max Planck Institute for the Science of Light, G\"unter-Scharowsky-Stra{\ss}e 1/Bau 24, 91058 Erlangen, Germany\\
$^2$ Friedrich-Alexander University of Erlangen-N\"urnberg Erlangen Centre for Astroparticle Physics Erwin-Rommel-Str. 1 91058 Erlangen\\
$^3$ Experimental Astrophysics Group, Space Sciences Laboratory, University of California Berkeley, CA 94720, USA \\
$^4$ European organization for Nuclear Research, CERN, CH-1211, Geneve 23, Switzerland\\
$^5$ Department of Physics, M. V. Lomonosov Moscow State University, Leninskie Gory, 119991 Moscow, Russia\\
$^6$ University of Erlangen-N\"urnberg, Staudtstra{\ss}e 7/B2, 91058 Erlangen, Germany
}

\email{felix.just@mpl.mpg.de} 



\begin{abstract}
During the last decades, multi-pixel detectors have been developed capable of registering single photons. The newly developed Hybrid Photon Detector camera has a remarkable property that it has not only spatial but also temporal resolution. In this work, we use this device for the detection of non-classical light from spontaneous parametric down-conversion and use two-photon correlations for the absolute calibration of its quantum efficiency.
\end{abstract}




\section{Introduction}
During the last three decades, several types of multi-pixel detectors have been developed, capable of registering single photons. These include intensified CCD (ICCD) cameras relying on amplification via micro channel plates, electron-multiplied CCD (EMCCD) cameras, and simpler devices, called multi-pixel photon counters, or solid-state photomultipliers. In the latter, there is no possibility of addressing particular pixels as the signals from different pixels are joined together at the output. As a result, due to the cross-talk between the pixels, false multi-photon correlations arise \cite{rech2008}, which complicate the extraction of useful information \cite{kalashnikov2011}. EMCCD and ICCD cameras are free from this drawback as they enable addressing particular pixels and, although they also manifest cross-talk, one can eliminate its influence by combining pixels into `clusters' or choosing well separated pixels. With such cameras, despite their relatively low quantum efficiencies, photon counting is a standard technique \cite{edgar2012} and ICCD and EMCCD cameras are traditionally used for imaging single-photon emitters such as molecules \cite{moerner2003}. They are also applied to the measurement of intensity correlation functions for two-photon light emitted via spontaneous parametric down-conversion (SPDC) \cite{hamar2010} and ghost imaging based on this effect \cite{tasca2013}. Using SPDC radiation, the quantum efficiency of an ICCD camera was calibrated \cite{perina2012} by means of a fit to the measured correlation functions. Recently, an ICCD camera was also used for spatially resolved characterization of single-photon emitters and their clusters \cite{shcherbina2014}. A huge disadvantage of ICCD cameras (and even more in the case of EMCCDs) for correlation measurements and therefore coincident detection is the necessity for a long measurement time. Even though ICCD cameras can be gated within time windows as short as a few ns, usually a whole frame can reveal only a single coincidence event. Furthermore, the readout times of such devices are usually much longer than the gating which additionally slows down the data acquisition significantly.

The new hybrid photon detector (HPD) can measure not only the position of an incident photon but also its time of arrival. Each individual pixel of the camera matrix is capable of tracking the point in time at which it was triggered by a photon. Even though it can only do so once per frame in the present configuration, many correlation measurement schemes based on multi-pixel detectors might be significantly improved by using this device. In contrast to other detectors of this kind, each pixel can measure events individually at the same time. Furthermore, one can even go one step further and think of more sophisticated correlation measurement schemes in the future, which might lift the once-per-frame constraint entirely. In this work we perform the first coincidence measurements with such a detector. We apply this camera to the detection of non-classical light from SPDC and use two-photon correlations for the absolute calibration of its quantum efficiency.

\section{The Hybrid Photon Detector}
\label{sec:HPD}

\begin{figure}[tb]
\centering
\includegraphics[width=0.5\textwidth]{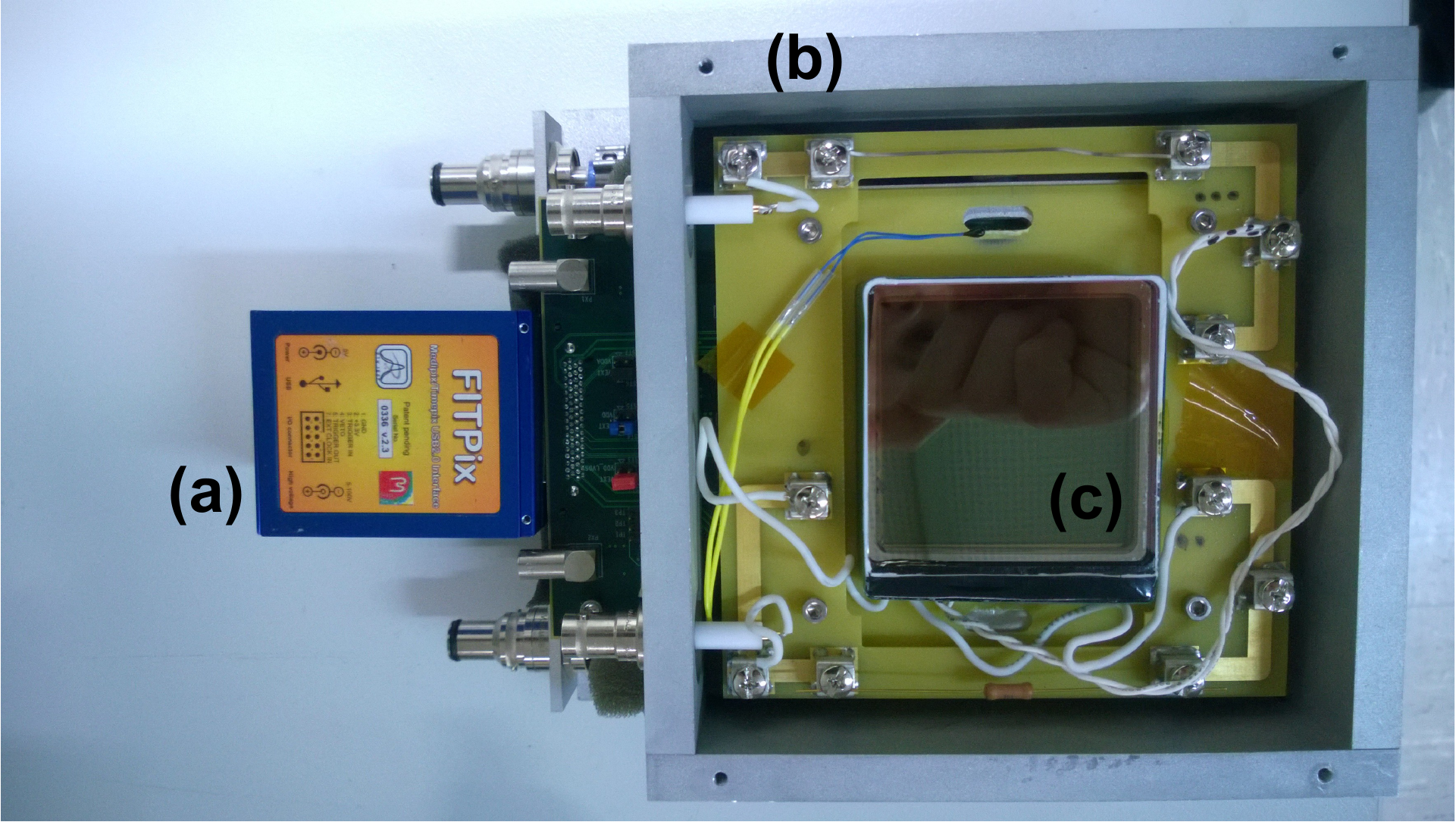}
\caption{A photograph of the hybrid photon detector. (a) read-out electronics, (b) PCB board with housing, (c) the tube.}
\label{photo_HPD}
\end{figure}

The Hybrid Photon Detector (HPD) is a first-of-its-kind device which provides a high temporal and spatial resolution for the detection of single photons in the optical range. The proof-of-principle experiments were carried out by J. Vallerga et al. at the Space-Science-Laboratory in Berkeley \cite{vallerga2008}. It was further developed by the Medipix collaboration and fabricated by Photonis Inc.\footnote{Photonis USA Inc., 6170 Research Road, Suite 208, TX-75033 Frisco, USA}. The detector control and read-out (FitPix) was developed at the IAEP in Prague \cite{kraus2011}. For the data acquisition the Pixelman software \cite{holy2006} was conceived. Depending on the mode of operation the spatial resolution can be as good as 6 \textmu m; the timing resolution can be $\approx$ 10 ns for a single pixel. A photograph is shown in Fig. \ref{photo_HPD}. Details about the HPD can be found in \cite{vallerga2014}.


\begin{figure}[tb]
\centering
\includegraphics[width=0.8\textwidth]{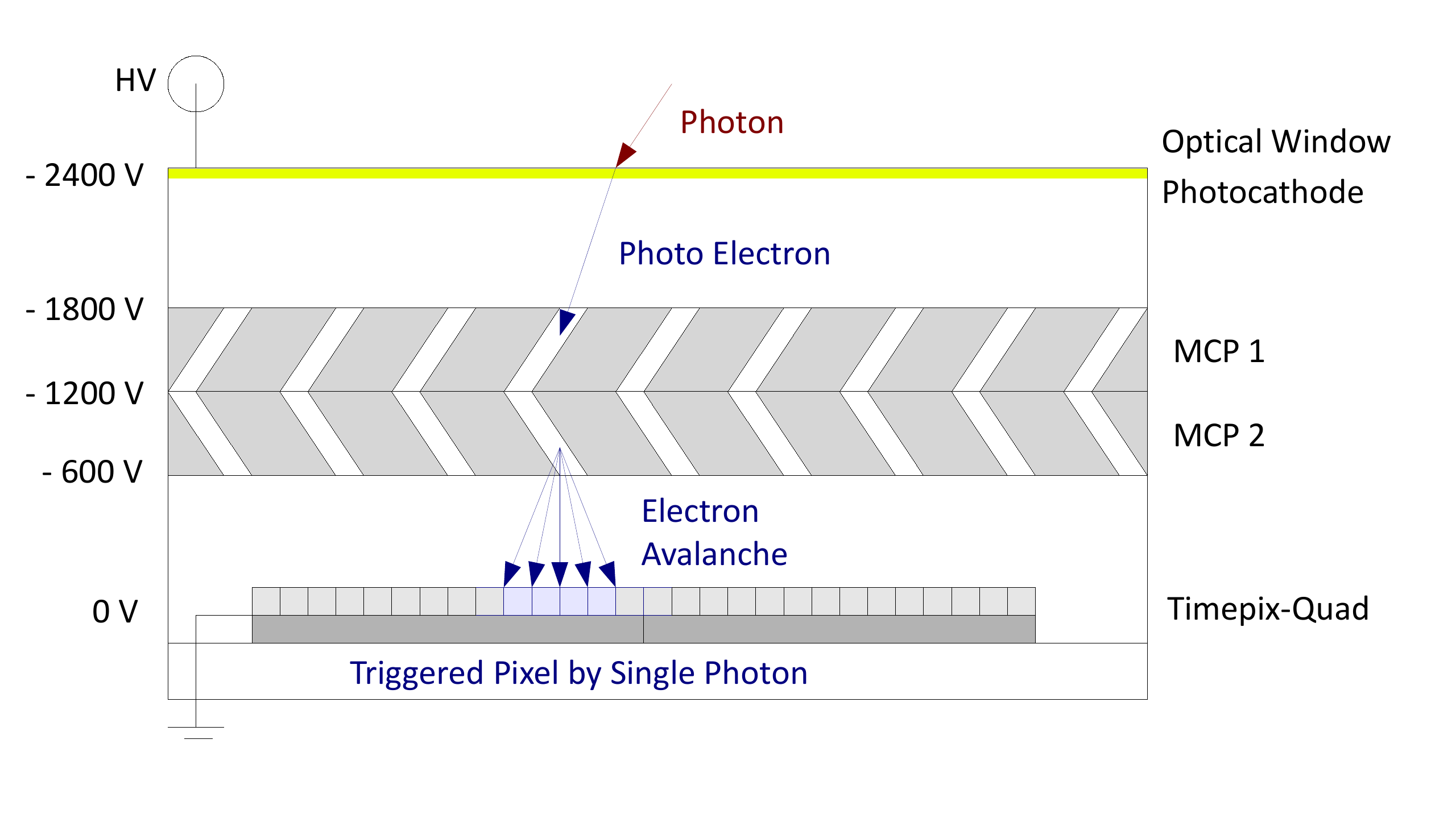}
\caption{A schematic cut through the HPD tube.}
\label{fig:scheme_HPD}
\end{figure}

\begin{figure}[tb]
\centering
\includegraphics[width=0.5\textwidth]{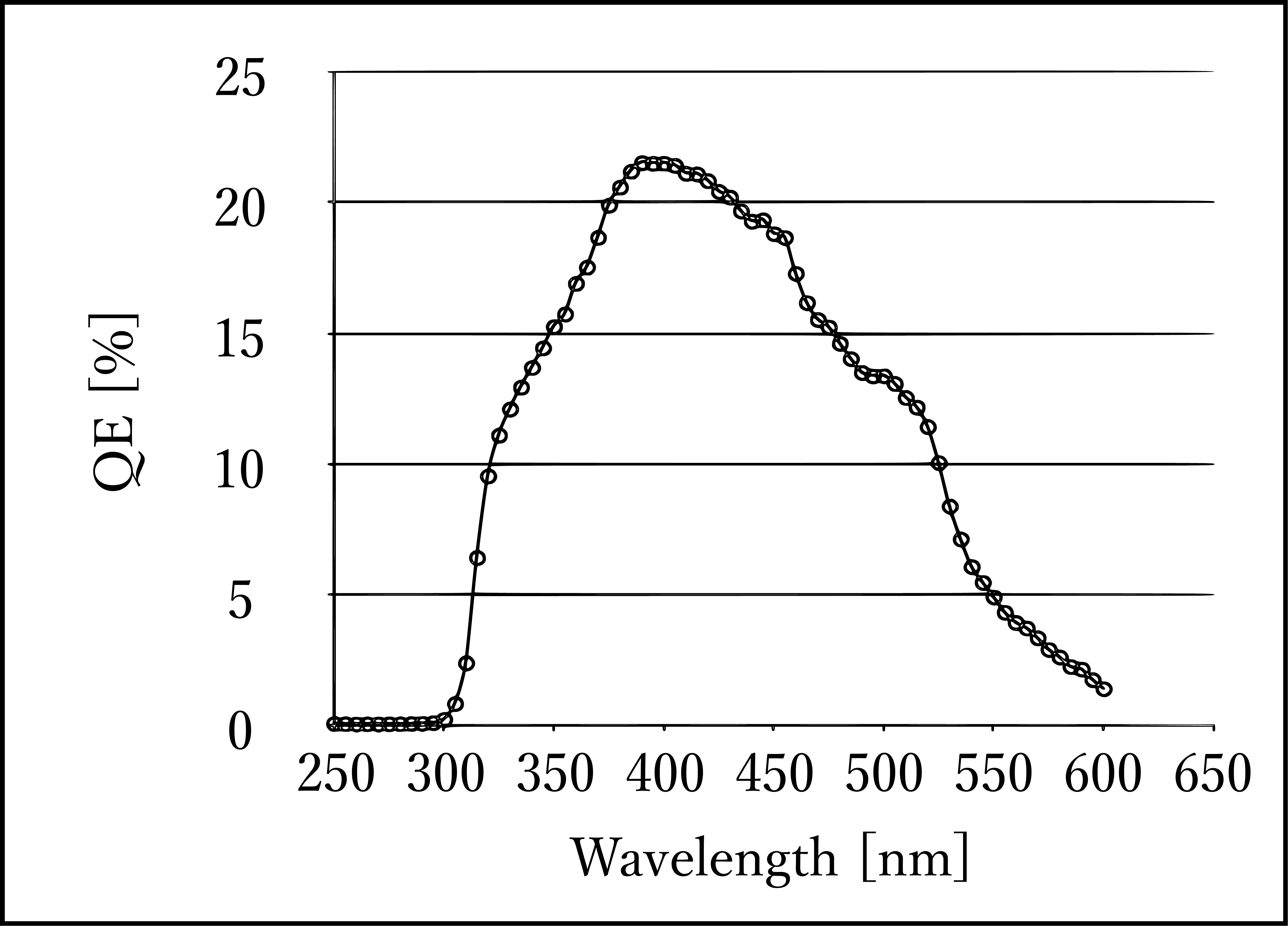}
\caption{Quantum efficiency of the photo cathode (measured by Photonis Inc.).}
\label{fig:qePhotoCath}
\end{figure}

The HPD consists of the read-out electronics (a), a printed circuit board (PCB) with housing (b) and the tube (c). The structure and functionality of the HPD tube is illustrated in Fig. \ref{fig:scheme_HPD}. At the top is an optically transparent window coated with a bi-alkali photocathode. Located 4 mm beneath the window are two microchannel-plates (MCPs) stacked onto each other. Underneath the MCPs, four Timepix application specific circuits (ASICs) \cite{{llopart2009}} are located in a 2x2 layout. During regular operation the cathode is at nominal voltage of -2.4 kV. The top surface of the upper MCP is at -1.8 kV. Both MCPs have a voltage gradient of 600 V across them. Hence, the bottom surface of the lower MCP is at -600 V. The Timepix-ASIC is grounded and the tube is at a vacuum pressure of about $10^{-10}\textrm{ mbar}$.

At the photocathode, an optical photon is absorbed and a photo-electron is emitted. Due to the voltage difference the photo-electron is pulled towards the MCP. An MCP is a thin piece of glass with co-aligned pores with a typical diameter of 6 - 20 \textmu m. A photo-electron entering the pore will produce multiple secondary electrons due to its kinetic energy when it hits the wall of the pore. This process happens many times within the pore which leads to electron multiplication. From one single photo-electron about $10^6$ electrons are produced in this process by the two consecutive MCPs. This avalanche of secondary electrons is collected on the input electrodes of the pixels of the Timepix-ASIC.\par
\begin{figure}[tb]
\centering
\includegraphics[width=0.8\textwidth]{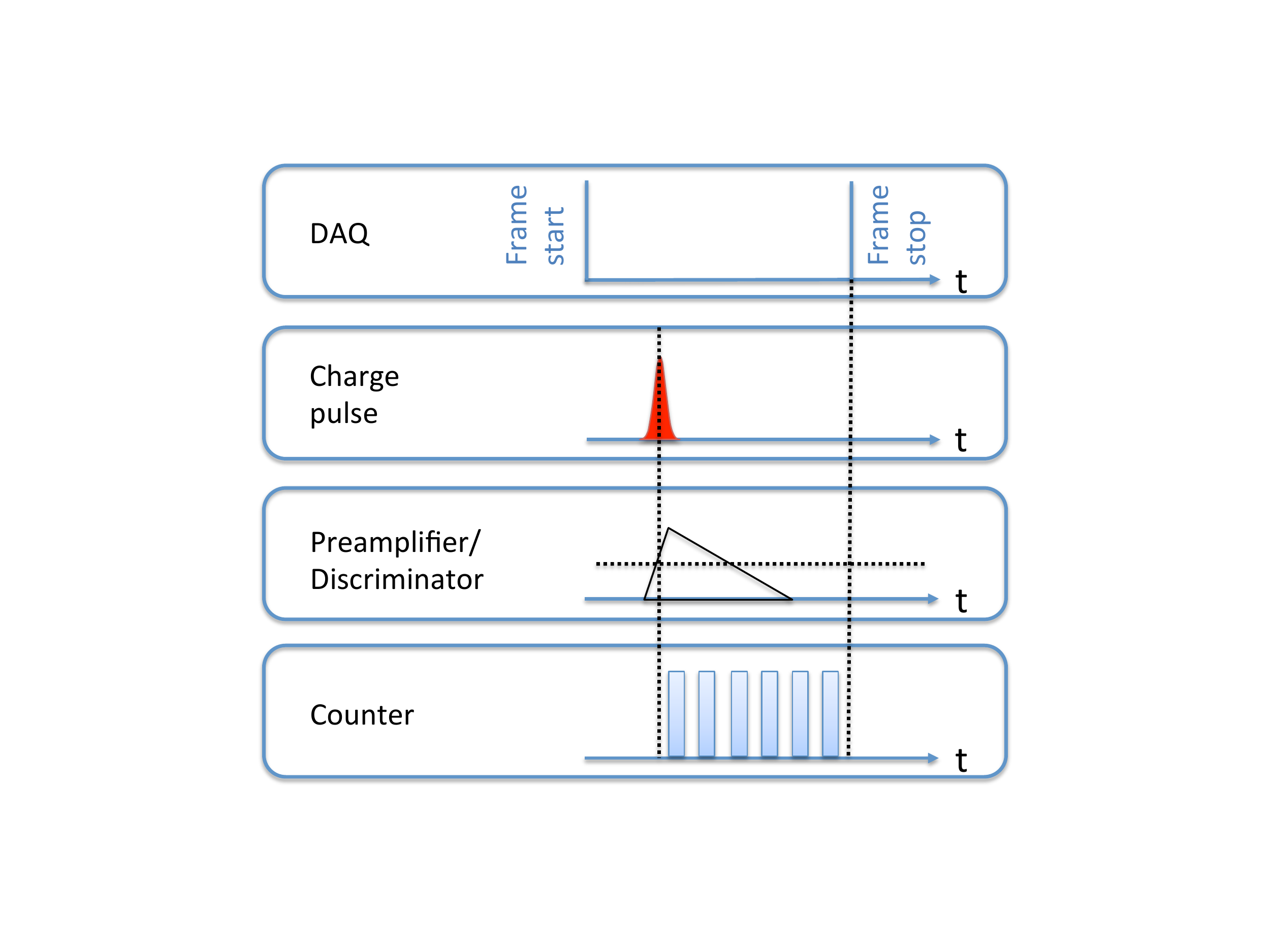}
\caption{The operation of the Timepix in the ``time-of-arrival'' (ToA) mode. After the charge pulse is amplified and converted to a voltage pulse, it is discriminated against a threshold level. The time-over-arrival is the number of clock counts between the first time the voltage pulse rises over the threshold and the end of the frame and thus the number of clock counts until the frame ends.}
\label{fig:toa_mode}
\end{figure}

The Timepix-ASIC has 256 x 256 pixel with a pixel size of 55 \textmu m. Every pixel contains its own amplifier and logical unit which coverts the collected charge into a voltage pulse and also includes a counter. The voltage pulse is discriminated against a global threshold in every pixel. The threshold corresponds to about 1000 electrons. The Timepix can be operated in three different modes. In the `time-over-threshold' mode, the Timepix returns how much charge was collected in every pixel by the number of digital clock pulses that are counted during the time when the voltage pulse is over the threshold. The second way the camera can be operated in is the counting mode in which each event above the detection threshold increases the counter of a hit pixel. Finally, in the `time-of-arrival' mode (ToA) the detector provides a timing information for every pixel by the amount of digital clock pulses from the first time the voltage pulse goes over the threshold until the frame ends (Fig. \ref{fig:toa_mode}). Since we are interested in correlation measurements the only mode of operation considered in this paper will be the ToA mode.

\section{Absolute calibration}
In conventional calibration methods, the quantum efficiency of the detector under test (DUT) is obtained via comparison with a reference detector. The non-classical state of light emitted by SPDC~\cite{louisell61} allows for a reference free, or absolute, method to calibrate the quantum efficiency of a photon counting detector \cite{ware2004}.
Such a scheme was proposed as a method for calibration by Klyshko \cite{klyshko1980} (and hence is often referred to as the \textit{Klyshko} method), and later implemented in experiment \cite{malygin1981_1, malygin1981_2}. It should be noted, however, that the first measurement of quantum efficiency using SPDC was performed by Burnham and Weinberg \cite{burnham70}. The idea is to utilise the photon-number correlations in SPDC. The Hamiltonian of the process is $\hat{H} \propto \int \mathrm{d} \mathbf{k}_s \mathrm{d} \mathbf{k}_i \hat{a}^{\dagger}_{\mathbf{k}_s} \hat{b}^{\dagger}_{\mathbf{k}_i} + \mathrm{h.c.}$, where $\hat{a}^{\dagger}_{\mathbf{k}_s}$ and $ \hat{b}^{\dagger}_{\mathbf{k}_i}$ are the creation operators for photons with wavevectors $\mathbf{k}_{s,i}$ in the signal ($s$) and idler ($i$) modes respectively. The quantum state of the emitted photons is obtained by applying the time evolution $| \psi (t) \rangle = \mathrm{e}^{-\frac{\imath}{\hbar} \hat{H} t} | 0 \rangle$ and expanding it to the first order only. In other words, we consider the low-gain regime where photons are exclusively created in pairs, $| \psi \rangle = |0 \rangle + c | 1 \rangle_{\mathbf{k}_s} | 1 \rangle_{\mathbf{k}_i}$, with $|c|\ll1$. Therefore, from detecting the signal photon, one can infer with absolute certainty that a corresponding photon has to be present in the idler arm.

Let us assume the DUT is exposed to the signal radiation while a second detector, from here on referred to as trigger detector, is placed in the idler arm of an SPDC source. The numbers of counts during a certain time interval are
\begin{align}
\label{eqn:qe1}
N_{\mathrm{DUT}} &= \eta_{\mathrm{DUT}} N, \nonumber \\
N_{\mathrm{trigger}} &= \eta_{\mathrm{trigger}} N,
\end{align}
where $\eta_{\mathrm{DUT}}$ and $\eta_{\mathrm{trigger}}$ are the quantum efficiencies of the two detectors and $N$ is the number of photon pairs emitted by the source in the said time interval. Accordingly, the number of coincidence events is
\begin{equation}
N_{\mathrm{coinc}} = \eta_{\mathrm{DUT}} \eta_{\mathrm{trigger}} N.
\end{equation}
For the quantum efficiency of the DUT it follows directly that
\begin{equation}
\label{eqn:qe2}
\eta_{\mathrm{DUT}} = \frac{N_{\mathrm{coinc}}}{N_{\mathrm{trigger}}}.
\end{equation}
Using coincidence measurements, we are thus able to ascertain $\eta_{\mathrm{DUT}}$ independently of the quantum efficiency of an imperfect trigger detector or any reference whatsoever.

One crucial remark to this method however is that equation (\ref{eqn:qe2}) does not strictly describe the quantum efficiency  of the DUT alone, but rather the complete DUT-arm of the setup, i.e., the whole channel starting from the crystal until the detection including any elements in between which could cause losses of any kind \cite{ware2004}. In practice that means one should include the lowest amount of optical elements in the DUT channel possible. Fortunately, one of the advantages of the scheme utilising an SPDC source is that the spectral selection can be performed in the trigger channel and thus will not affect the measurement.

Another issue arises from the fact that SPCD photons are emitted at a certain angular width. The spread of the emission angle is due to the transverse wave vector components of the pump, the width of the phasematching function (which is given by the length of the nonlinear crystal), and the selected frequency bandwidth \cite{ware2004}.  In order to avoid events in the trigger channel which have no corresponding counterpart in the DUT and hence would lead to an artificial decrease of $\eta_{\mathrm{DUT}}$ in equation (\ref{eqn:qe2}), one has to make sure that the angular acceptance of the DUT is greater than that of the trigger detector. Similar arguments of course, hold for the spectral width.\par
Finally one also has to take into account that there are always some accidental coincidences, which originate from the overlapping of photon pairs and have to be subtracted from the total number of coincidence events. Consequently we used a modified equation (\ref{eqn:qe2}) to obtain the actual quantum efficiency under laboratory conditions \cite{kwiat1994}:
\begin{equation}
\label{eqn:qe3}
\eta_{\mathrm{DUT}} = \frac{N_{\mathrm{coinc}} - N_{\mathrm{acc}}}{N_{\mathrm{trigger}} - N_{\mathrm{dark}}}.
\end{equation}
Here $N_{\mathrm{acc}}$ denotes the number of accidental events and $N_{\mathrm{dark}}$ are the dark counts of the trigger detector during the measurement time.
\section{Setup}

\subsection{Optical setup}
\label{sec:optSetup}
\begin{figure}[tb]
\centering
\includegraphics[width=0.8\textwidth]{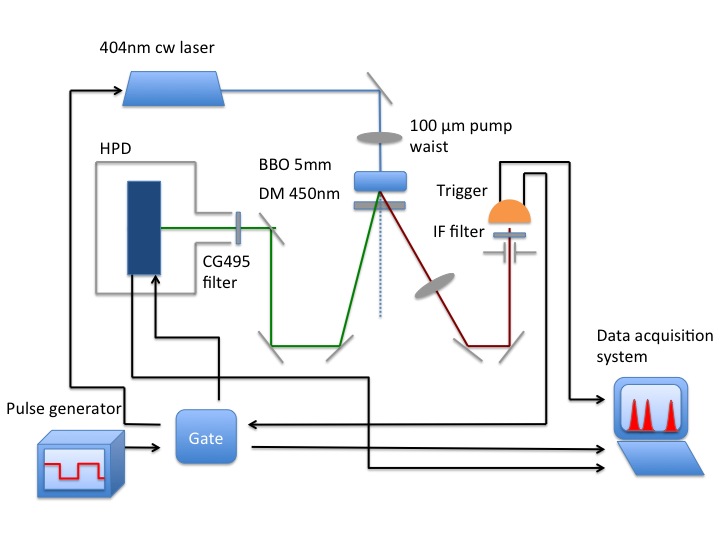}
\caption{Optical setup.}
\label{fig:setup}
\end{figure}

A sketch of the optical setup is given in Fig. \ref{fig:setup}. In order to create the photon pairs, a $\beta$-Barium Borate (BBO) crystal of 5mm length was pumped by a cw diode laser operating at a wavelength of 405nm. The pump power was about 1.5mW and the beam was focused down to a waist of about 100$\upmu$m. At first, the phasematching was chosen such that the the photons in the camera channel were emitted at 560nm and under an angle of $4^\circ$, while the trigger photons were emitted at $10^\circ$ with a wavelength of 1463nm. To ensure detection of correlated photon pairs a bandpass filter of 12nm width was placed in the trigger channel. The wavelengths could be tuned by replacing the filter and adjusting the angle between the optic axis of the crystal and the pump beam accordingly to preserve the alignment. As previously mentioned, it is necessary that the trigger detector is exposed to either a more limited or to exactly the same angular spectrum corresponding to the spectrum in the DUT channel. An iris placed in the far field ensures this condition to be fulfilled. Due to the rather large sensitive area of the detector, we placed the HPD in a box to shield it from ambient light. The trigger photons were then fibre coupled to a single mode fibre and detected by a Scontel superconducting nano-wire detector.

In the last section we stated that for the calibration it is beneficial to do all filtering in the trigger channel, since losses there will not affect the quantum efficiency of the DUT.  Nevertheless a dichroic mirror (DM450) and a colour glass filter (CG495) had to be included in the DUT arm. Even though the pump is spatially separated from the signal, fluorescence and scattered light from the pump originating from the crystal could easily saturate the camera because their spectrum almost coincides with the maximum sensitivity range of the HPD photocathode.

\subsection{Electronics and data acquisition}
\label{sec:DAQ}
To obtain the quantum efficiency from equation (\ref{eqn:qe2}), the number of coincidences between the trigger detector and the DUT has to be determined. To achieve this the following scheme was realized.

\begin{figure}[tb]
\centering
\subfloat[]{\includegraphics[width=0.4\textwidth,height=12em]{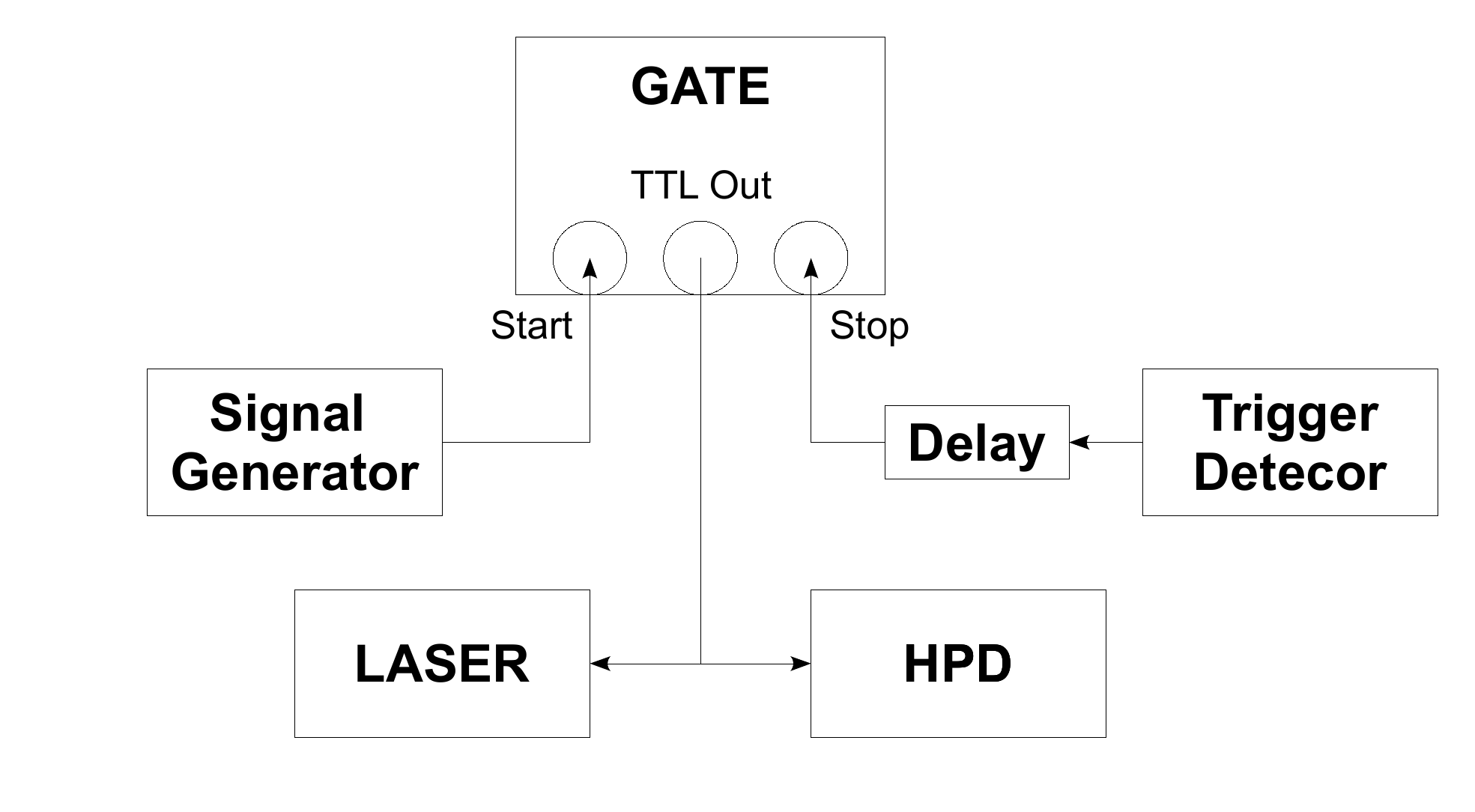}}
\subfloat[]{\includegraphics[width=0.55\textwidth]{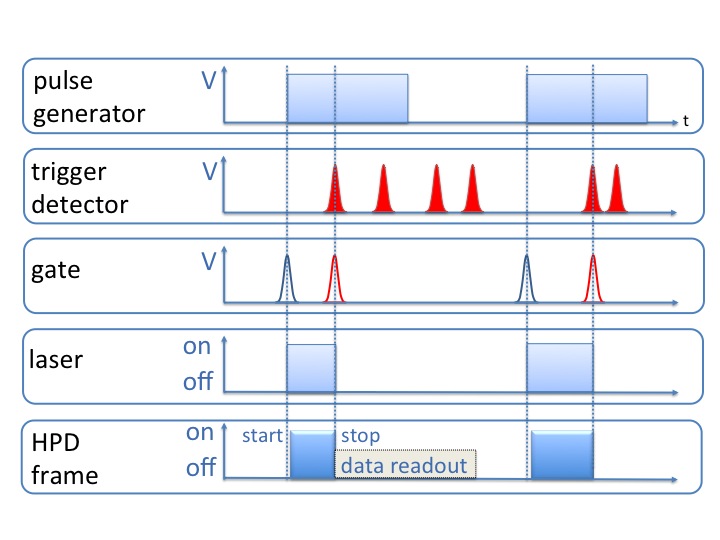}}
\caption{(a) Diagram circuit for the coincidence measurement. (b) Scheme of coincidence readout}
\label{fig:coincScheme}
\end{figure}

The HPD delivers the data not continuously but in frames. Therefore, we synchronized the laser and the HPD data acquisition by a frequency generator and a gate to reduce both the dark noise and the overall photon flux on the camera. A scheme of the electronics is shown in Fig. \ref{fig:coincScheme}.
The frequency generator delivered the start signal to the gate at a fixed rate. Its output signal (TTL) was then split into the start signal for the HPD and the start signal for the laser. 
The stop signal on the gate came from the trigger detector. We discriminated the output of the trigger detector and then sent a TTL signal to the gate. This signal was delayed by 1.2 \textmu s. Upon the arrival of the stop signal, both the frame and the laser were stopped. After that the data was read out from the Timepix chips ($\approx 200$ ms).  As in the ToA mode the time until the frame end was obtained for each pixel individually, we saw coincident events between both detectors at a particular ToA, corresponding to the delay between the trigger detector and the frame end of the frame in the HPD. The circuit diagram and the logical scheme are depicted in Fig. \ref{fig:coincScheme}.

\section{Data analysis and results}
\subsection{Data analysis}
\label{sec:dataAnalysis}
\begin{figure}[tb]
\centering
 \subfloat[Single Frame]{\includegraphics[width=0.3\textwidth]{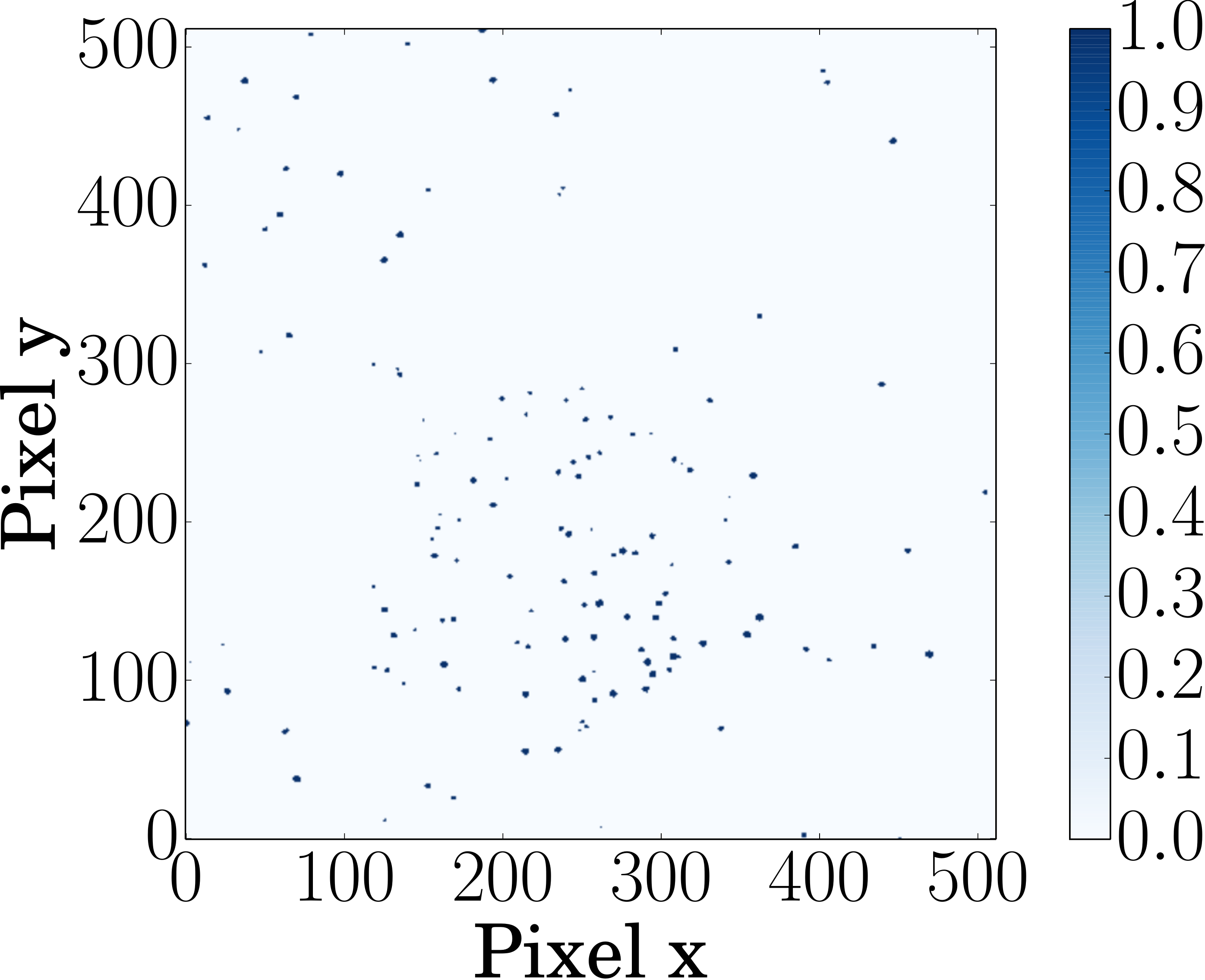} \label{fig:single_frame}}\hspace{10pt}
\subfloat[All Frames]{\includegraphics[width=0.3\textwidth]{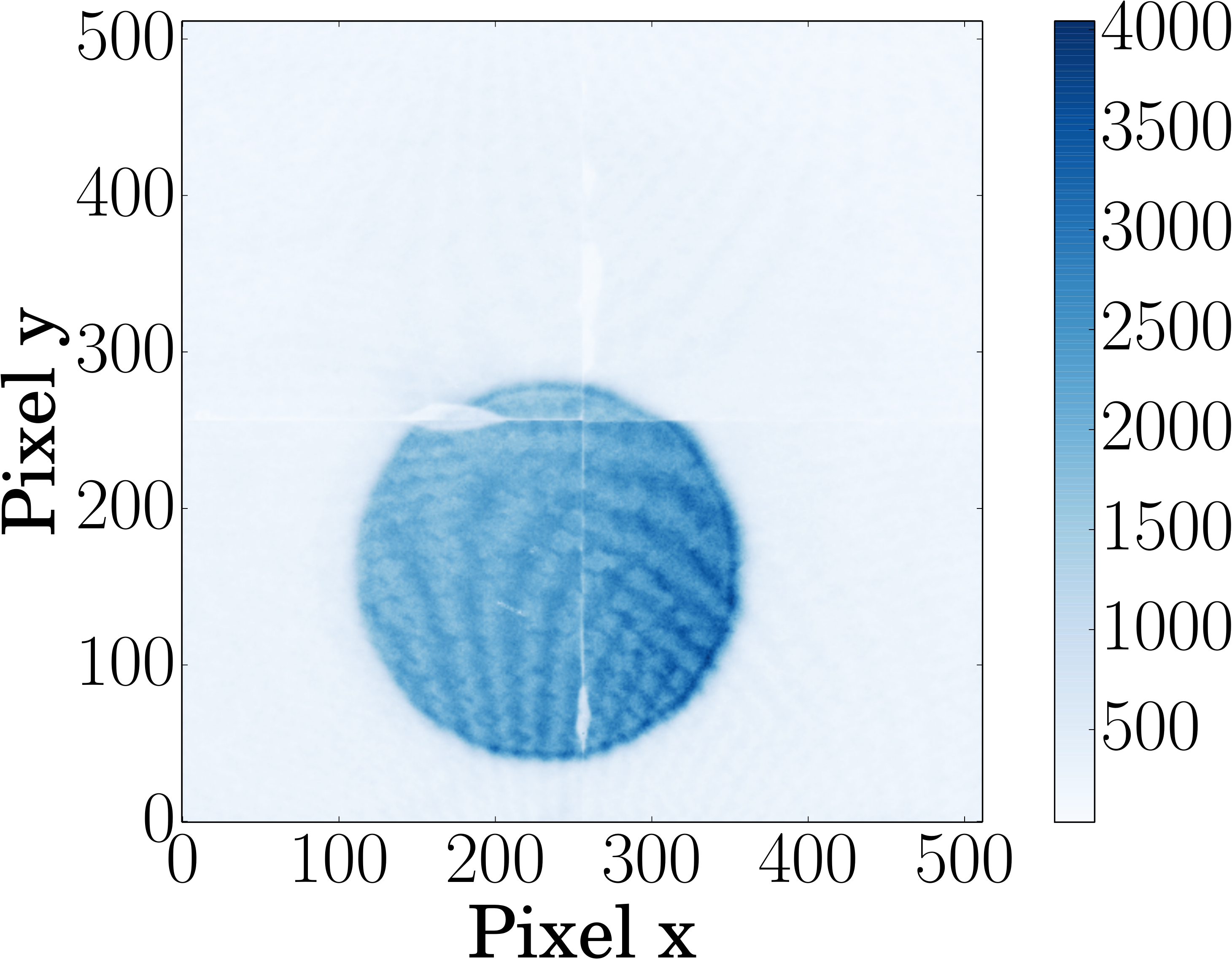}  \label{fig:int_frame}}\hspace{10pt}
\subfloat[Coincident Events]{\includegraphics[width=0.3\textwidth]{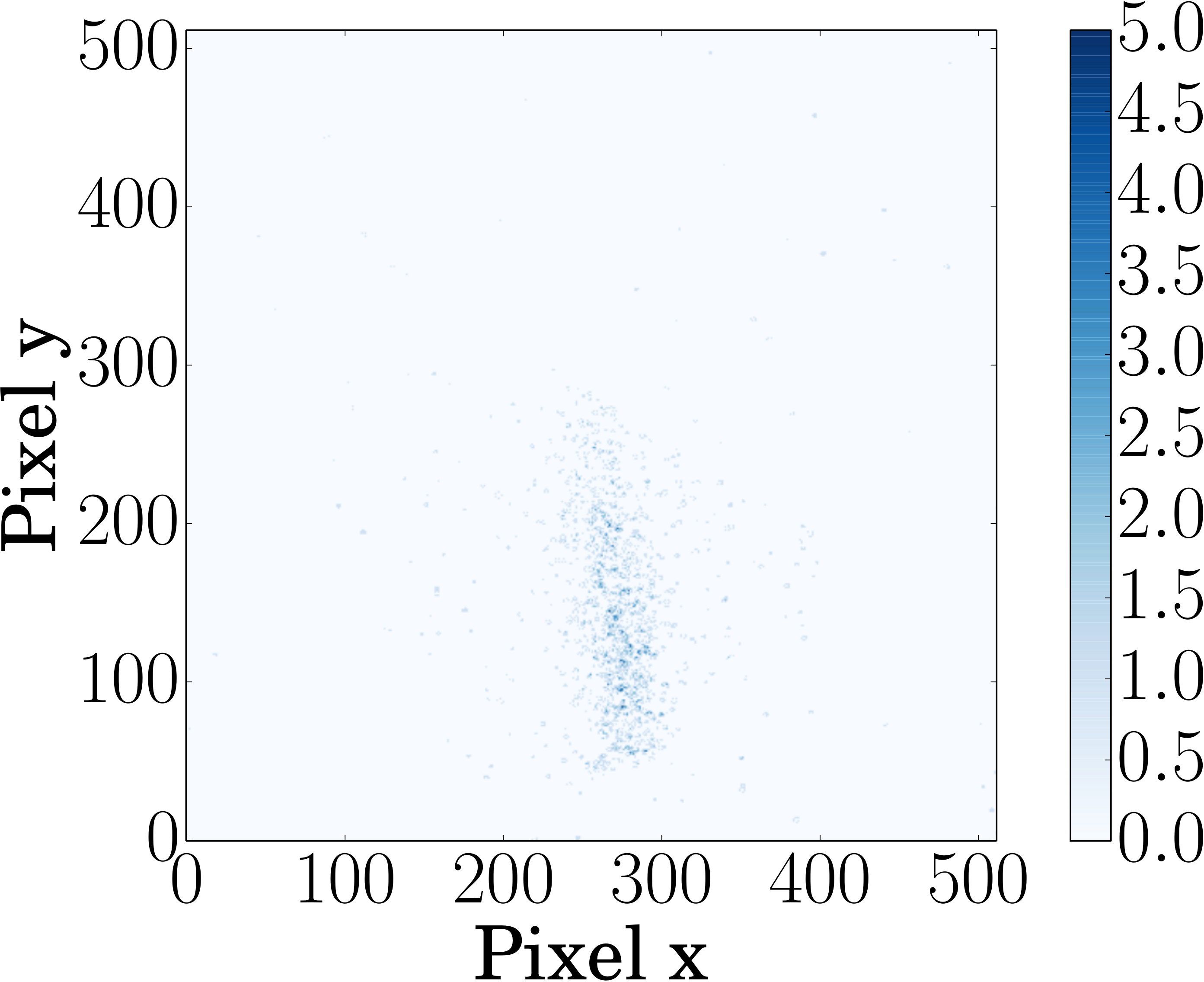}  \label{fig:coin_frame}}
\caption{\protect\subref{fig:single_frame} Examples of a single frame, \protect\subref{fig:int_frame} the result of the integration over all frames for one measurement cycle, and \protect\subref{fig:coin_frame} the post-selected events corresponding to the coincidence peak. The colour bar indicates the number of events.}
\end{figure}

\begin{figure}[tb]
\centering
\subfloat[ToA distribution of coincidences]
{
\label{fig:toa_spec}
\includegraphics[width=0.5\textwidth]{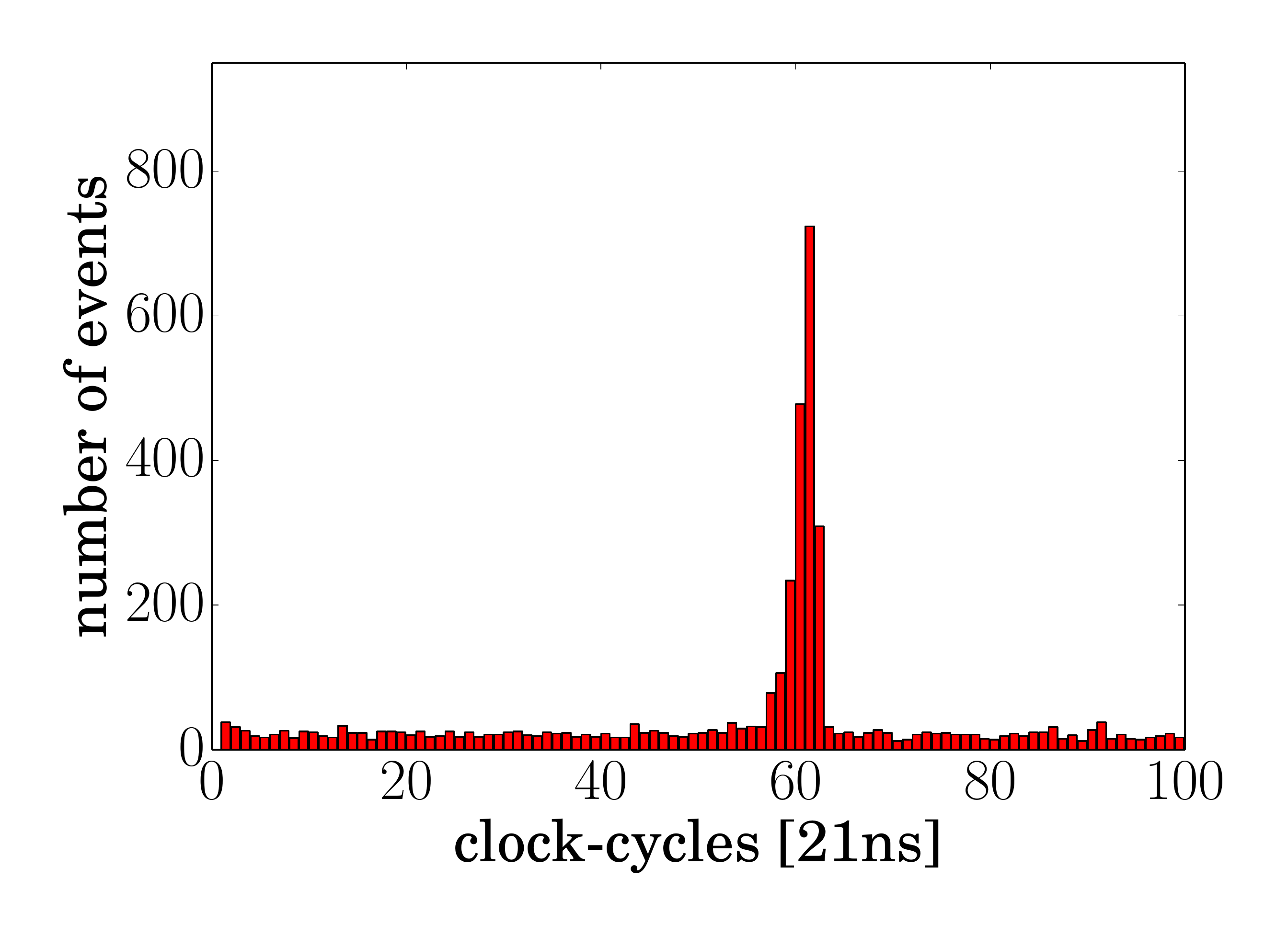}
}
\subfloat[Acquisition time histogram]
{
\label{fig:ac_time_dist}
\includegraphics[width=0.5\textwidth]{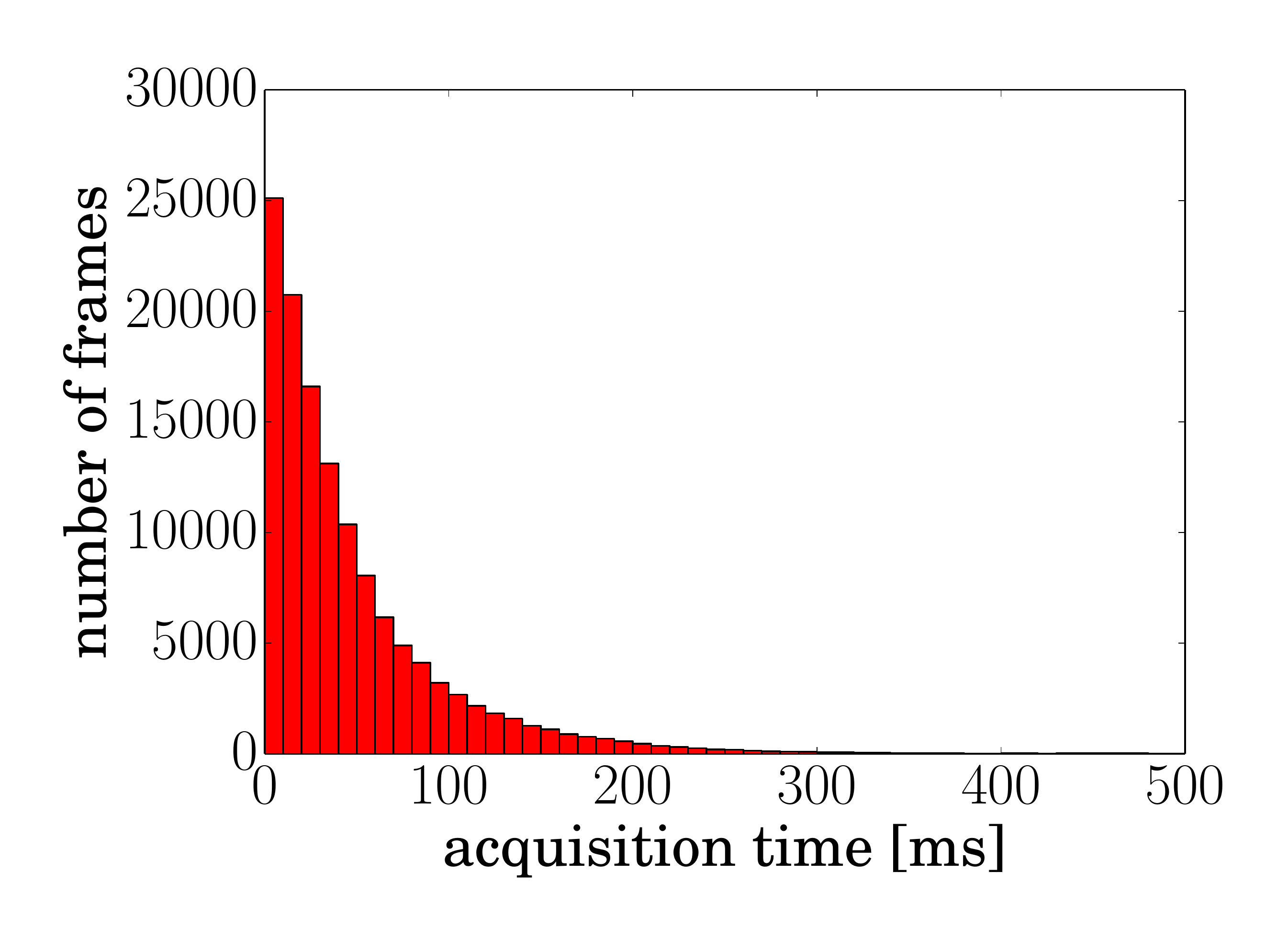}
}
\caption{\protect\subref{fig:toa_spec} Example of an integrated ToA spectrum of one complete measurement cycle for a single wavelength. The coincident peak appears at 20 clock cycles (=1275 ns). \protect\subref{fig:ac_time_dist} Example of a typical acquisition time histogram.}
\end{figure}

Fig \subref*{fig:single_frame} shows the spatial distribution of detected photon events for one typical frame. Every single photon event is detected as a cluster of multiple triggered pixels ($\approx$ 5 on average) on the matrix. To obtain the timing information for a cluster we averaged over the time-stamps (ToA) of its pixels. The number of coincidences between both detectors during a complete measurement cycle was obtained by the integration over all frames. Accidental coincidence events manifest themselves as a background in the histogram. The time distribution of coincidence events (number of clusters plotted versus ToA) for a single measurement is shown in Fig. \subref*{fig:toa_spec}. One can clearly distinguish the peak of coincident events from the background of accidental coincidences. The peak is centered at about 1275 ns which corresponds to the delay that we introduced into the system. Please note that the abscissae are in clock cycles of 21 ns each (corresponding to a ToA clock cycle frequency of 48 MHz) until the end of the frame. Therefore, events with larger ToA values are actually earlier than events close to the origin (the frame end).


An image of all events integrated over the complete measurement is given in Fig. \subref*{fig:int_frame}. Here we observe the whole SPDC spectrum (all possible phasematching processes). Its circular shape is governed by the aperture of the box that was used to protect the HPD from ambient light. Fig. \subref*{fig:coin_frame} shows only events coincident with the trigger. The line that is visible is actually part of an arc of the SPDC radiation (usually emitted along a cone) corresponding to the wavelength range selected by the filter in the trigger arm.

The level of accidental coincidences, given by the background of the coincidence distribution, was determined by taking the average of the distribution, excluding its peak. Since the trigger detector provides a rate in counts per second while from the coincidence time distribution the number of counts is obtained, we had to read-out the total acquisition time for the HPD. In Fig.\subref*{fig:ac_time_dist} the distribution of frame-times for a complete run is shown. The integral of this distribution was used as the total acquisition time. The dark noise of the trigger detector was recorded during the time intervals when the laser was off. 

\subsection{Results}
\label{sec:results}
\begin{figure}[tb]
\centering
\includegraphics[width=0.8\textwidth]{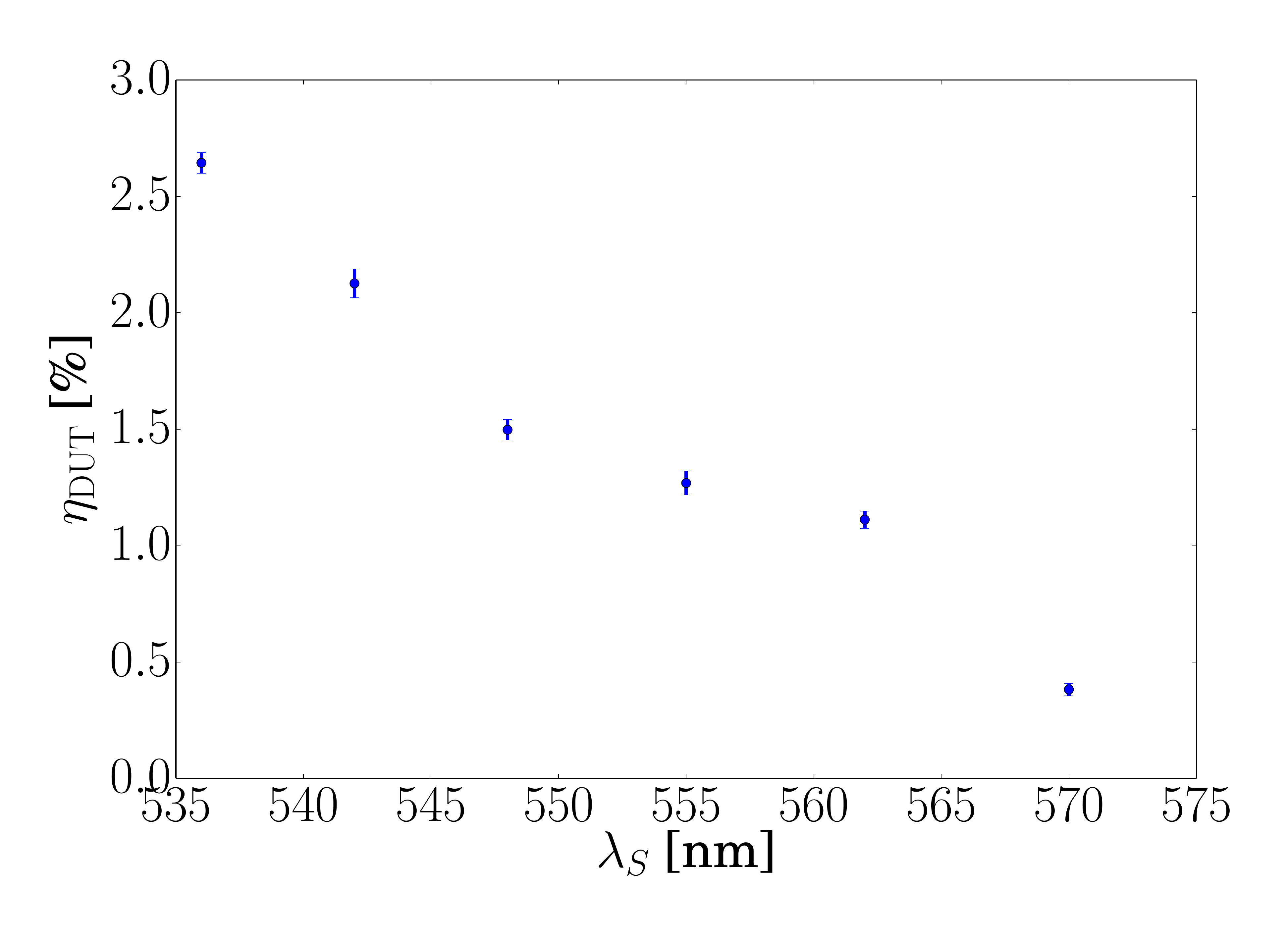}
\caption{Quantum efficiency as a function of the wavelength at 2400 V MCP stack voltage.}
\label{fig:qePlot}
\end{figure}

Figure \ref{fig:qePlot} shows the quantum efficiency measured as a function of the wavelength.  As expected (see Fig. \ref{fig:qePhotoCath}), the quantum efficiency increases towards shorter wavelengths. We found the error bars by assuming a Poisson distribution for the count rates and performing the usual Gaussian error propagation. The main contribution to the errors stems from the fluctuations in the dark counts of the trigger detector. Those were assumed to be Poissonian distributed as well. However during some of the measurements we observed a minor systematic drift of the dark count rate and hence the distribution might have non-random contributions which in turn could lead to a slight underestimation of errors.
Compared to the quantum efficiency of the photocathode, the absolute quantum efficiency of the HPD is about a factor of 2 lower. The main reason for the discrepancy between the quantum efficiencies of the photocathode and the overall system is due to the ratio between the area of the holes and the non-sensitive parts of the MCP. If a photo-electron does not fall into an MCP hole but onto the non-sensitive area, it is not detected. The geometry of the MCP is a hexogonal lattice of 25 \textmu m round pores with the pore centers separated by 32 \textmu m. From this we obtain  a probability of 0.55 for an electron to hit a pore. This corresponds to the ratio between the results of the absolute calibration and the calibration of the photocathode only. Notwithstanding the factor of 2, within the spectral range of our measurements, the quantum efficiency curve follows quite nicely that of the specifications of the photocathode. \par
\begin{figure}[tb]
\centering
\includegraphics[width=0.8\textwidth]{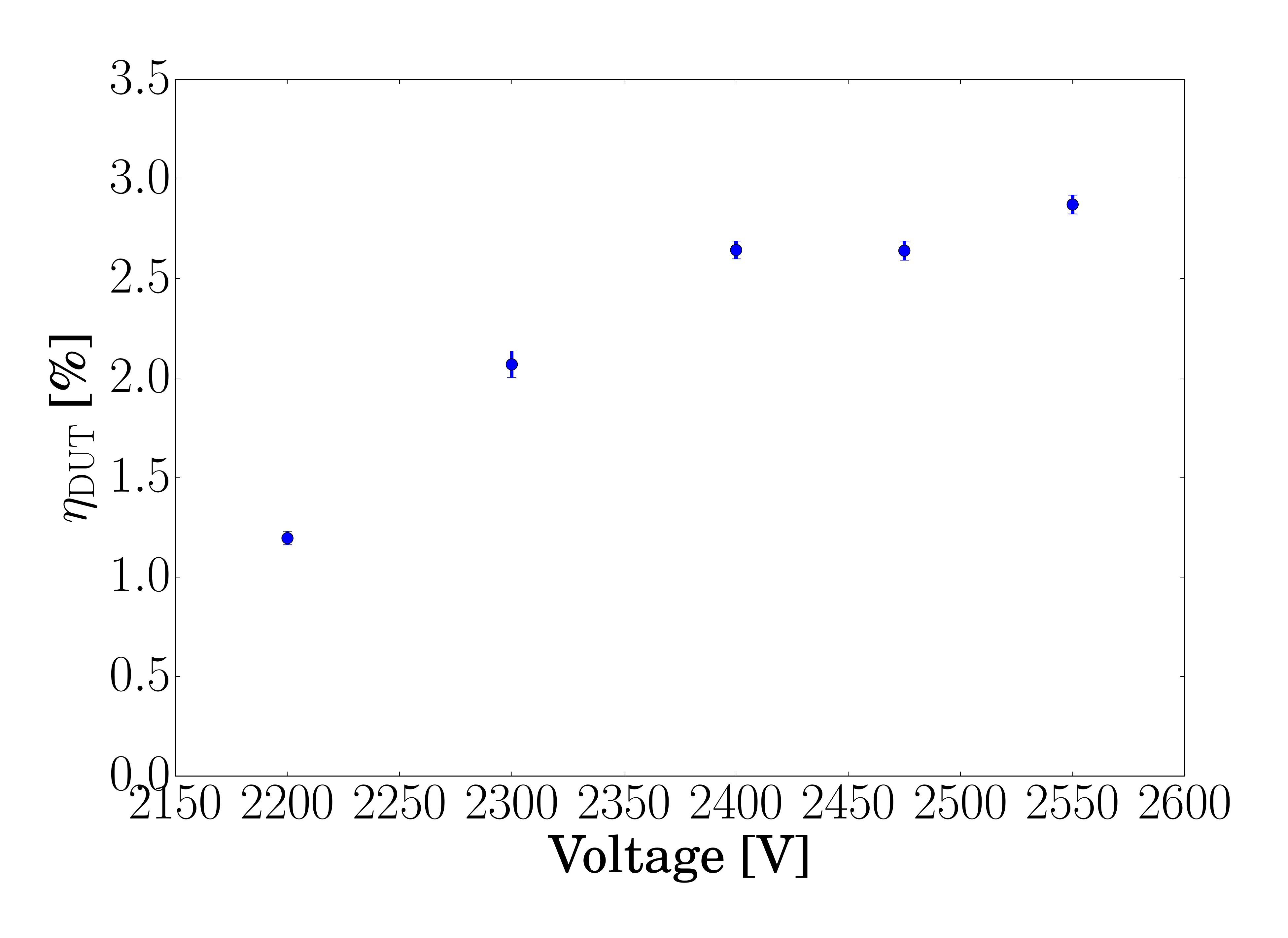}
\caption{Quantum efficiency as a function of the MCP stack voltage at 536 nm..}
\label{fig:qePlot_voltage}
\end{figure}
One parameter which directly affects the quantum efficiency of the HPD is the voltage applied to the tube, therefore we also measured $\eta_{\mathrm{DUT}}$ at a fixed wavelength of 536 nm as a function of the voltage. The amplification in the MCP depends directly on the kinetic energy of the electrons involved. At higher voltages the electrons in the pores will experience a stronger acceleration and thus create a bigger avalanche which in turn is more likely to be detected by the Timepix chips. These considerations make it apparent that the MCP gain and thus the quantum efficiency of the tube increases with the voltage. The dependence of $\eta$ on the voltage is illustrated in Fig. \ref{fig:qePlot_voltage}. We can see a saturation of the quantum efficiency at high voltages. The reason behind this is that at sufficiently high voltages (and thus amplifications) every electron emitted by the photocathode will result in at least one detection event. The results can be found in agreement with the previous measurements  \cite{vallerga2014}. In Table \ref{tab:res} the results for the calibration at different wavelengths are summarised. Please note that these values are subject to losses due to the absorption in the dichroic mirror and the colour glass (both $\approx10\%$). Column three of Table \ref{tab:res} gives the quantum efficiency corrected for the filter transmissions.\par
\begin{table}[tb]
\centering
\begin{tabular}{| c || c | c | c |}
\hline
wavelength [nm] & $\eta_{\mathrm{DUT}}$ [\%] & $\eta^{\mathrm{corrected}}_{\mathrm{DUT}}$ [\%] & $\delta \eta_{\mathrm{DUT}}$ [\%] \\
\hline \hline
536 & 2.64 & 3.19 &  0.04\\
542 & 2.13 & 2.59 & 0.06\\
548 & 1.50 & 2.05 & 0.04\\
555 & 1.27 & 1.68 & 0.05\\
562 & 1.11 & 1.36 & 0.04\\
570 & 0.38 & 0.48 & 0.03\\
\hline
\end{tabular}
\caption{Results of the absolute calibration measurements}
\label{tab:res}
\end{table}

\section{Conclusion}
We successfully implemented the first measurements on nonclassical light with the HPD operating in the single-photon regime. An SPDC source has been used to perform the absolute calibration measurement of the HPD from 535 nm to 570 nm. Furthermore we investigated the effect of the voltage applied to the MCPs on the quantum efficiency.  Those first measurements demonstrate the HPD's capability for coincidence acquisition. This opens the door for correlation measurements in general and thus many of the standard applications in quantum optics. We believe that in the future, the HPD's unique capabilities namely the availability of both space and time resolution have the potential to improve the existing experiments or even allow new ones.
The main drawback is the long readout time of the chip and the resulting low frame rate. This however will be remedied with the next generation of the Timepix-ASIC \cite{gromov2011} which will provide not only higher temporal resolution but also continuous data output.
\section*{Acknowledgments}
We are very grateful to the Medipix collaboration for the development of the HPD. This work was supported in part by ERA-Net.RUS (project Nanoquint).

\end{document}